\documentclass[acmsmall]{acmart}
\usepackage{siunitx}  
\usepackage{algorithmic}
\usepackage{algorithm}

\newcommand{\e}{\mathrm{e}}
\newcommand{\sato}{}

\setcopyright{acmcopyright}
\copyrightyear{0000}
\acmYear{0000}
\acmDOI{XXXXXXX.XXXXXXX}

\acmJournal{TECS}
\acmVolume{00}
\acmNumber{0}
\acmArticle{000}
\acmMonth{0}

\acmSubmissionID{64}

\begin{document}

\title{Fast Parameter Optimization of Delayed Feedback Reservoir with Backpropagation and Gradient Descent}

\author{Sosei Ikeda}
\affiliation{%
  \institution{Kyoto University}
  \city{Kyoto}
  \country{Japan}}
\email{sikeda@easter.kuee.kyoto-u.ac.jp}

\author{Hiromitsu Awano}
\affiliation{%
  \institution{Kyoto University}
  \city{Kyoto}
  \country{Japan}} 
\email{awano@i.kyoto-u.ac.jp}

\author{Takashi Sato}
\affiliation{%
  \institution{Kyoto University}
  \city{Kyoto}
  \country{Japan}}
\email{takashi@i.kyoto-u.ac.jp}

\renewcommand{\shortauthors}{Ikeda et al.}

\begin{abstract}
%
A delayed feedback reservoir (DFR) is a reservoir computing system well-suited for hardware implementations.
However, achieving high accuracy in DFRs depends heavily on selecting appropriate hyperparameters.
Conventionally, due to the presence of a non-linear circuit block in the DFR,
the grid search has only been the preferred method, which is computationally intensive and time-consuming, and thus performed offline.
This paper presents a fast and accurate parameter optimization method for DFRs.
To this end, we leverage the well-known backpropagation and gradient descent framework with the state-of-the-art DFR model
for the first time
to facilitate parameter optimization.
To achieve the highest accuracy with reduced memory usage,
we further propose truncated backpropagation strategy applicable to the recursive dot-product reservoir representation. 
With the proposed lightweight implementation, 
the computation time has been significantly reduced by up to 1/700 of grid search.


\end{abstract}



\received{00 January 0000}
\received[revised]{00 January 0000}
\received[accepted]{00 January 0000}
\maketitle


\section{Introduction}\label{sec:intro}

Reservoir computing (RC) is a machine learning method closely associated with the neuromorphic concept~\cite{tanaka2019recent}.
It uses a reservoir, which nonlinearly transforms inputs into a high-dimensional space.
The reservoir weights are fixed in RC; only the output layer weights are trained~\cite{lukovsevivcius2009reservoir}. 

A delayed feedback reservoir (DFR)~\cite{appeltant2011information} is a specific type of RC system.
It consists of a nonlinear element and a feedback loop.
The nonlinear element often adopts a neuromorphic Mackey--Glass equation, which models cells in the blood~\cite{mackey1977oscillation}.
The structural simplicity of DFR makes it an attractive candidate for embedded hardware implementation, with low power consumption being a notable advantage among RCs~\cite{tanaka2019recent}.
Various implementation methods for RC have been proposed, including electronic circuits~\cite{appeltant2011information}, optical elements~\cite{larger2012photonic}, etc.

The performance of RC heavily depends on the reservoir weights~\cite{tanaka2019recent}.
In the case of DFR, grid search has been commonly used to optimize the weights or parameters~\cite{appeltant2011information}.
However, grid search faces a significant scalability problem.
In the case of DFR, both the regularization parameter of the linear regression at the output layer
and the reservoir parameters must be optimized as hyperparameters,
resulting in a high--dimensional search space.
Additionally, this optimization process can be highly nonlinear, necessitating a finely divided grid,
leading to a significant increase in
computation time as the number of optimizing parameters grows.
In the context of embedded applications,
online learning is crucial as it aligns well with the low-power implementation that DFR enables.
However, the challenges associated with training DFR hinder its realization
in online learning scenarios.
To overcome these obstacles and achieve online learning in DFR,
fast learning methods are essential.

Backpropagation and gradient descent~\cite{amari1993backpropagation} are widely used and proven effective
training methods for neural networks. However,
to the best of the authors' knowledge, no prior study has applied these methods to DFRs.
One of the reasons for this absense is that DFRs use a nonlinear element often based on physical phenomena,
making it challenging to 
backpropagate residues, or requiring complex calculations~\cite{nakajima2022physical}.
%
%

To address this limitation, the authors propose a parameter optimization method for DFRs based on backpropagation and gradient descent,
leveraging the concept of modular DFR~\cite{ikeda2023modular}.
The modular DFR model reduces the number of parameters in the reservoir layer and divides the nonlinear element into multiple blocks, facilitating the application of backpropagation.
With this approach, the authors demonstrate that fast learning is achievable for DFRs.
As only the regularization parameter of linear regression in the output layer is treated as a hyperparameter, and
recursively defined reservoir state of the DFR, which serves as a feature for the output layer, is approximated, the fixed number of epochs are required.
The important contributions of this study can be summarized as follows:
\begin{enumerate}
 \setlength{\itemsep}{0pt}
 \item Proposal of a method to optimize reservoir parameters using backpropagation and gradient descent
       based on the modular DFR model that allows for selecting a nonlinear function whose derivatives can be
       efficiently obtained while maintaining accuracy.
 \item Simplification of the backpropagation
       calculation by reducing the number of reservoir
       features used retrospectively. This enhancement leads
       to faster computation and can reduce memory usage
       required for backpropagation.
       
 \item Demonstration of the faster learning of the proposed
       method using multiple classification tasks.  The method achieves
       comparable accuracy to traditional grid search
       with magnitude, up to 700x, reduced computation time.

\end{enumerate}

\section{DFR}\label{sec:dfr}


\subsection{Concept of DFR}\label{subsec:dfr}

The DFR is a specific implementation of RC, utilizing a nonlinear element combined with a delayed feedback loop.
RC has a structure that reflects past input, making it suitable for processing time series~\cite{tanaka2019recent}.
This approach proves to be more straightforward for hardware implementation than other RC schemes due to 
its simplistic structure~\cite{tanaka2019recent}.
Typically, hardware realizations of DFRs employ analog circuits in the reservoir layer~\cite{appeltant2011information,soriano2014delay}.

\begin{figure}[tb]
 \centering
 \includegraphics[width=1.0\linewidth]{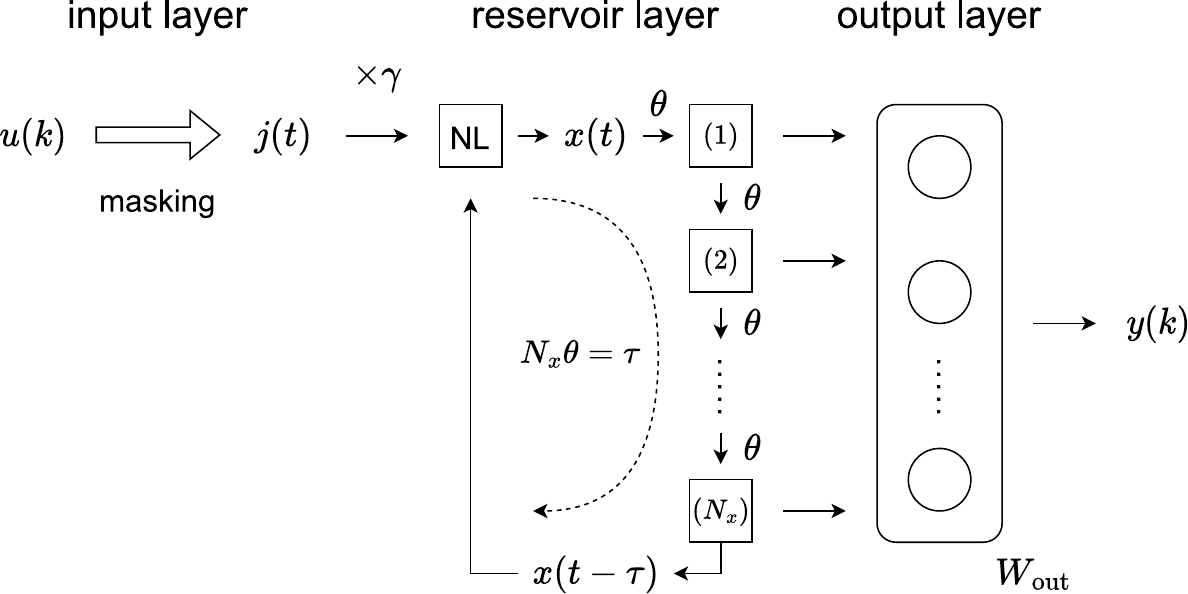}
 \caption{Conceptual diagram of DFR\@.
 The reservoir consists of a nonlinear element (NL) and a feedback loop with a total delay $\tau$.
 The feedback loop comprises $N_x$ virtual nodes with a time interval $\theta$.
 }\label{fig:dfr}
 \vspace{-5mm}
\end{figure}

To demonstrate the operation and the challenges in applying efficient learning methods,
we explain the standard analog implementation of DFR\@.
Fig.~\ref{fig:dfr} illustrates its conceptual diagram.
Initially, the digital time series input, $u(k)$, sampled at a period $\tau$, is converted to an analog signal, $i(t)$.
This signal, $i(t)$, undergoes masking wherein it is multiplied by the mask signal with a faster sample rate $\theta$.
The resulting signal $j(t)$ is subsequently scaled by $\gamma$, added to the reservoir feedback, and fed into
the nonlinear element (NL) to generate the output, $x(t)$.
The NL is mathematically represented by the following delay differential equation:
\begin{align}
 \frac{\mathrm d}{{\mathrm d}t}x(t)=F(x(t-\tau), \gamma j(t)).
 \label{eq:delay}
\end{align}
The Mackey--Glass (MG) model~\cite{mackey1977oscillation} is a commonly used NL in various studies~\cite{appeltant2011information,soriano2014delay,alomar2015digital}.
Its operation is expressed by the following equations:
\begin{align}
 \frac{\mathrm d}{{\mathrm d}t}x(t)=-x(t)+\eta f(x(t-\tau),\gamma j(t)), \label{eq:MG} \\
 f(x(t-\tau),\gamma j(t))=\frac{[x(t-\tau)+\gamma j(t)]}{1+{[x(t-\tau)+\gamma j(t)]}^p}. \label{eq:f}
\end{align}
Here, $p$ is an adjustable parameter.
In this context, the virtual nodes are connected at a time interval $\theta$ on
the feedback loop with a total delay of $\tau$.
The signal values at these virtual nodes serve as the features and are collected as a vector
of $N_x$ elements, referred to as the reservoir state, where $N_x\theta =\tau$.

The reservoir state is defined as follows:
\begin{align}
 \boldsymbol{x}(k)\equiv[x(k\tau -\theta ), x(k\tau -2\theta ), \ldots , x(k\tau -\tau)]. \label{eq:x}
\end{align}
The output is subsequently obtained by applying a linear transformation.

To enhance the ease of implementation, fully digital versions of DFRs have been proposed~\cite{appeltant2011information,alomar2015digital}.
In these digital DFRs, the signal, $\boldsymbol{j}(k)$ after the masking process is expressed as $\boldsymbol{j}(k)=\boldsymbol{m}u(k)$
where $\boldsymbol{m}$ is a mask vector of length $N_x$
and its elements are determined randomly and digitally~\cite{appeltant2011information}.

After masking, the nonlinear element and feedback loop of the DFR determine the reservoir state, $\boldsymbol{x}(k)$.
Assuming that $f$ in Eq.~(\ref{eq:MG}) is constant for a short time $\theta$, which is the time interval between the virtual nodes,
the differential equation in Eq.~(\ref{eq:MG}) is solved for $x(t)$ as follows:
\begin{align}
 x(t)=x_0\e^{-t}+\eta (1-\e^{-t})f(x(t-\tau)+\gamma j(t)),
\end{align}
where $x_0$ is the initial value of $x(t)$ at each $\theta$~\cite{appeltant2011information}.
The value of the next virtual node can be expressed as $x(\theta)$.
Assuming that the components of $\boldsymbol{x}(k)$ and $\boldsymbol{j}(k)$ are
\begin{align}
 \boldsymbol{x}(k)&\equiv[x(k)_1,x(k)_2,\ldots,x(k)_{N_x}],\\
 \boldsymbol{j}(k)&\equiv[j(k)_1,j(k)_2,\ldots,j(k)_{N_x}],
\end{align}
$\boldsymbol{x}(k)$ is obtained recurrently as follows:
\begin{align}
 x(k)_n = x(k)_{n-1}\e^{-\theta}+(1-\e^{-\theta})f(x(k-1)_n, j(k)_n).
\end{align}
Here, the initial value of the reservoir state is set to 0.
The construction and operation of the output layer are the same as in an analog implementation.

In either implementations, the recursive nature of the network
construction and the differentially defined NL make grid search the only
effective method for parameter optimization.

\subsection{DPRR}\label{subsec:dprr}

In classification tasks, a single output is expected for multiple time-series inputs.
The number of features obtained by the reservoir layer depends on the length of the input series,
necessitating the features converted into an intermediate representation of fixed length
called the {\em reservoir representation}~\cite{ikeda2022hardware}.
This conversion allows the features to be processed by an output layer of a fixed size~\cite{chen2013model}.

Numerous reservoir representations have been proposed~\cite{pascanu2015malware,appeltant2011information,cabessa2021efficient,chen2013model,bianchi2020reservoir}.
Currently, DPRR is considered the best in terms of both accuracy and circuit size~\cite{ikeda2022hardware}.
Let the elements of $\boldsymbol{x}(k)$ (i.e., the features) be $[x(k)_1, x(k)_2, \ldots, x(k)_{N_x}]$.
The vector of a node in the reservoir:
\begin{align}
 \boldsymbol{x}_n\equiv[x(1)_n, x(2)_n, \ldots, x(T)_n],
\end{align}
can be considered to be the time evolution of the node state.
Here, $T$ denotes the length of a time series.
Taking the dot product of these vectors for two nodes yields:
\begin{align}
 \boldsymbol{x}_i\cdot\boldsymbol{x}_j^- = \sum_{k=1}^{T}x(k)_ix(k-1)_j \;(i, j=1, 2, \ldots, N_x).
 \label{eq:dprr1}
\end{align}
In addition, the reservoir state itself is used as a feature:
\begin{align}
 \boldsymbol{x}_i\cdot\textbf{1} = \sum_{k=1}^{T}x(k)_i \;(i=1, 2, \ldots, N_x).
 \label{eq:dprr2}
\end{align} 
The DPRR is defined as a vector $\boldsymbol{r}$ by concatenating Eqs.~(\ref{eq:dprr1}) and (\ref{eq:dprr2}), which consists of $N_x\cdot (N_x+1)$ values, i.e.,
$\boldsymbol{r} \equiv \mbox{vec}\left(\sum^T_{i=1} \boldsymbol{x}(k) \boldsymbol{x'}(k-1)\right)$, where $\boldsymbol{x'}(k-1) = [\boldsymbol{x}(k), 1]$.
The output is obtained using the output layer $W_\mathrm{out}$, where $N_r$ represents the number of features in a reservoir representation:
\begin{align}
 \boldsymbol{y}=W_{\mathrm{out}}\boldsymbol{r}+\boldsymbol{b}=\sum_{n=1}^{N_r}\boldsymbol{w}_n\boldsymbol{r}_n+\boldsymbol{b}.
 \label{eq:output}
\end{align}
For each of the components of $r$, multiple reservoir state features are used in the derivation, making backpropagation complicated and
difficult.

\subsection{Modular DFR}\label{subsec:modular}


The challenge in designing a DFR lies in determining the design of the NL\@.
In analog implementations utilize elements obey a delay differential equation.
In digital implementations, the goal is often to replicate the behavior of elements used in existing analog implementations.

\begin{figure}[tb]
 \centering
 \includegraphics[width=0.8\linewidth]{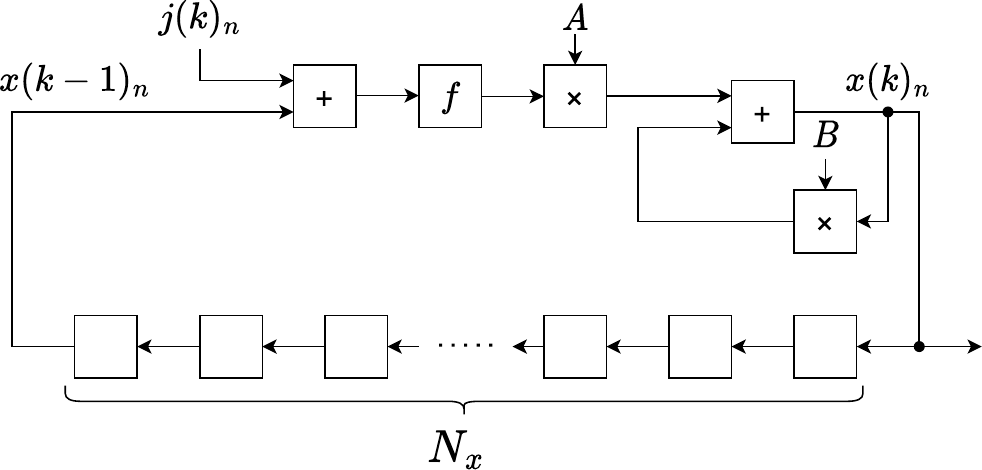}
 \caption{
 Block diagram of reservoir processing in the modular DFR model.
 The block labeled ``$f$'' operates as a one-input, one-output function $f$ in Eq.~(\ref{eq:MG}).
 Only two parameters, $A$ and $B$, have to be optimized.
 }\label{fig:model_new}
 \vspace{-5mm}
\end{figure}

The modular DFR presents a model that simplifies the design of the NL in the digital implementation of DFR~\cite{ikeda2023modular}.
First, the NL operation is decomposed into multiple blocks, simplifying
the design target from a delay differential equation to a one-input, one-output function $f$.
This reduction also decreases the number of parameters to be adjusted from 3 to 2 while preserving the same solution space.
Fig.~\ref{fig:model_new} illustrates the block diagram of the modular DFR model obtained.
Here, $x(k)_n$ can be represented as follows:
\begin{align}
x(k)_n=Af(j(k)_n+x(k-1)_n)+Bx(k)_{n-1}.
\label{eq:modular}
\end{align}
In Eq.~(\ref{eq:modular}), the parameters subject to optimization are $A$ and $B$.
In total, three parameters need to be optimized: these two and the regularization parameter for linear regression in the output layer.
However, using grid search for optimization is challenging due to the need to explore the three-dimensional parameter space, even after the number of parameters is reduced. Moreover,
improper selection of these three parameters could lead to insufficient accuracy.
Therefore, an efficient optimization strategy is essential to ensure effective parameter tuning.


\section{Parameter Optimization of DFR by backpropagation and gradient descent}\label{sec:propose}

In this section, we present the formulation of backpropagation for DFR\@.
We choose the modular DFR for its network structure and the task-specific extensibility of the NL\@.
The back propagation process is explained in reverse order, starting from the output layer, then the DPRR layer, and finally the reservoir layer, tracing back from the error in the output.
Special attention is given to the DPRR layer,
where back propagation occurs from multiple roots for a single value,
and the reservoir layer with a recursive structure.
Subsequently, we propose a truncation approach to the backpropagation to enable a more lightweight implementation.

\subsection{Backpropagation in the output layer}\label{subsec:bp_output}

\begin{figure}[tb]
 \centering
 \includegraphics[width=0.7\linewidth]{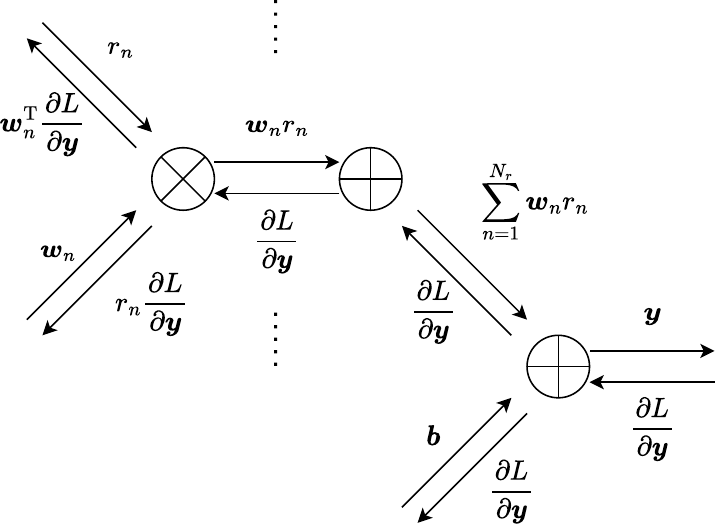}
 \caption{Computation graph of forward and backward propagation in the output layer.}\label{fig:bp_output}
 \vspace{-5mm}
\end{figure}

Fig.~\ref{fig:bp_output} shows the computation graph summarizing the backpropagation in the output layer.
First, the output $\boldsymbol{y}$ and the target output $\boldsymbol{d}$ are represented as the following vectors:
\begin{align}
\boldsymbol{y}=[y_1,y_2,\ldots,y_{N_y}]\ \mathrm{and}\ \boldsymbol{d}=[d_1,d_2,\ldots,d_{N_y}].
\end{align}
Here, $N_y$ represents the number of classes, and $\boldsymbol{d}$ is represented in a one-hot encoding.
The loss function is formulated using cross entropy error as:
\begin{align}
L=-\sum_{k=1}^{N_y}d_k\log y_k.
\end{align}
Then the partial derivative of $L$ with respect to $\boldsymbol{y}$ is given by:
\begin{align}
\frac{\partial L}{\partial \boldsymbol{y}}=\boldsymbol{y}-\boldsymbol{d}.
\end{align}
This derivative is propagated backwards through the network.

The computation in the output layer is expressed as Eq.~(\ref{eq:output}).
The partial derivatives of $L$ with respect to $\boldsymbol{b}$, $\boldsymbol{r}_n$, and $\boldsymbol{w}_n$ are respectively expressed as:
\begin{align}
\frac{\partial L}{\partial \boldsymbol{b}}=\frac{\partial L}{\partial \boldsymbol{y}}, \,
\frac{\partial L}{\partial \boldsymbol{r}_n}=\boldsymbol{w}_n^\mathrm{T}\frac{\partial L}{\partial \boldsymbol{y}}, \,
 \mathrm{and} \,
\frac{\partial L}{\partial \boldsymbol{w}_n}=\boldsymbol{r}_n\frac{\partial L}{\partial \boldsymbol{y}}.
\end{align}

\subsection{Backpropagation in the DPRR layer}
\label{subsec:bp_dprr}
\begin{figure}[tbh]
 \centering
 \includegraphics[width=0.9\linewidth]{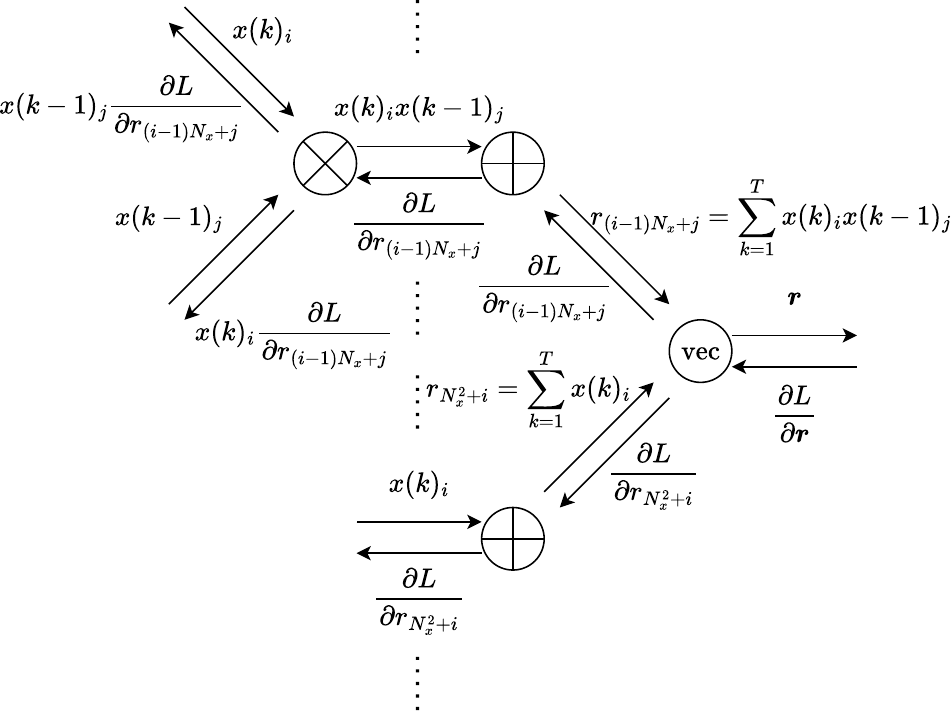}
 \caption{Computation graph of forward and backward propagation in the DPRR layer.
 }\label{fig:bp_dprr}
 \vspace{-5mm}
\end{figure}

Fig.~\ref{fig:bp_dprr} shows the computation graph summarizing the backpropagation in the DPRR layer.
First, each feature in the DPRR is defined as follows.
\begin{align}
r_{(i-1)N_x+j}&=\sum_{k=1}^{T}x(k)_ix(k-1)_j \;(i, j=1, \ldots, N_x),\\
r_{N_x^2+i}&=\sum_{k=1}^{T}x(k)_i \;(i=1, \ldots, N_x).
\end{align}
For $r_{(i-1)N_x+j}$, the following is backpropagated.
\begin{align}
\frac{\partial L}{\partial x(k)_i}&=x(k-1)_j\frac{\partial L}{\partial r_{(i-1)N_x+j}},\\
\frac{\partial L}{\partial x(k-1)_j}&=x(k)_i\frac{\partial L}{\partial r_{(i-1)N_x+j}}.
\end{align}
Subsequently, for $r_{N_x^2+i}$, the following is backpropagated.
\begin{align}
\frac{\partial L}{\partial x(k)_i}=\frac{\partial L}{\partial r_{N_x^2+i}}.
\end{align}
From the above, the following value (bpv) is backpropagated from the DPRR layer for the partial derivative of $L$ with respect to $x(k)_n$.
\begin{align}
(\mathrm{bpv})
 &=\sum_{j=1}^{N_x}x(k-1)_j\frac{\partial L}{\partial r_{(n-1)N_x+j}}\nonumber\\
& \quad +\sum_{i=1}^{N_x}x(k+1)_i\frac{\partial L}{\partial r_{(i-1)N_x+n}}+\frac{\partial L}{\partial r_{N_x^2+n}}.\label{eq:bpval}
\end{align}

\subsection{Backpropagation in the reservoir layer}\label{subsec:bp_reservoir}

\begin{figure}[tbh]
 \centering
 \includegraphics[width=1.0\linewidth]{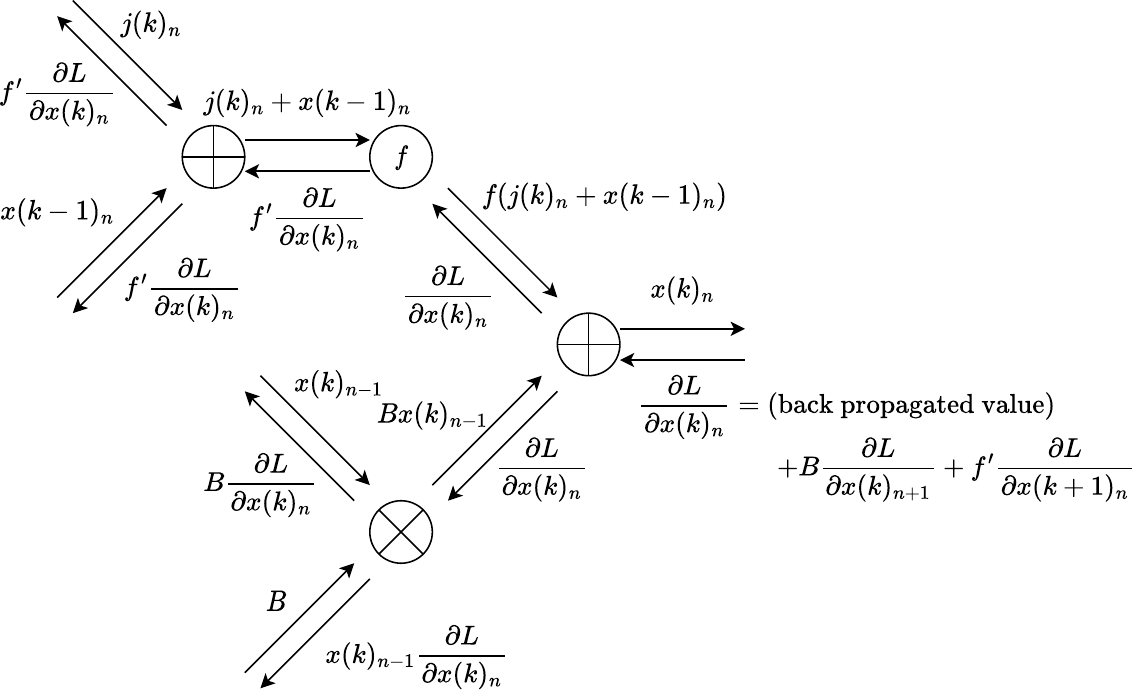}
 \caption{Computation graph of forward and backward propagation in the reservoir layer.
 }\label{fig:bp_reservoir}
 \vspace{-5mm} 
\end{figure}

Fig.~\ref{fig:bp_reservoir} shows the computation graph summarizing the backpropagation in the reservoir layer.
Note that $f$ has a constant multiplication parameter $A$.
First, the partial derivative of $L$ with respect to $Bx(k)_{n-1}$ is
\begin{align}
\frac{\partial L}{\partial (Bx(k)_{n-1})}=\frac{\partial L}{\partial x(k)_n}.
\end{align}
It is then expressed as follows.
\begin{align}
\frac{\partial L}{\partial x(k)_{n-1}}&=B\frac{\partial L}{\partial x(k)_n},\\
\frac{\partial L}{\partial B}&=x(k)_{n-1}\frac{\partial L}{\partial x(k)_n}.
\end{align}
Next, the partial derivative of $L$ with respect to $f(j(k)_n+x(k-1)_n)$ is
\begin{align}
\frac{\partial L}{\partial f(j(k)_n+x(k-1)_n)}=\frac{\partial L}{\partial x(k)_n}.
\end{align}
The partial derivative of $L$ with respect to the constant multiplication parameter $A$ of $f$ is
\begin{align}
\frac{\partial L}{\partial A}=\frac{\partial f(j(k)_n+x(k-1)_n)}{\partial A}\frac{\partial L}{\partial x(k)_n}.
\end{align}
Then, the following holds.
\begin{align}
\frac{\partial L}{\partial x(k-1)_n}&=\frac{\partial L}{\partial (j(k)_n+x(k-1)_n)}\nonumber\\
&=\frac{\partial f(j(k)_n+x(k-1)_n)}{\partial (j(k)_n+x(k-1)_n)}\frac{\partial L}{\partial x(k)_n}.
\end{align}
The partial derivative of $L$ with respect to $x(k)_n$ is recursively given as:
\begin{align}
\frac{\partial L}{\partial x(k)_n}&=(\mathrm{bpv})
+B\frac{\partial L}{\partial x(k)_{n+1}}+f'\frac{\partial L}{\partial x(k+1)_n}.\label{eq:Lx}
\end{align}
Therefore, the partial derivative of $L$ with respect to the constant multiplication parameter $A$ of $f$ and $B$ are respectively given by adding up for all past times as follows:
\begin{align}
\frac{\partial L}{\partial A}&=\sum_{k=1}^{T}\frac{\partial f(j(k)_n+x(k-1)_n)}{\partial A}\frac{\partial L}{\partial x(k)_n},\label{eq:LA}\\
\frac{\partial L}{\partial B}&=\sum_{k=1}^{T}x(k)_{n-1}\frac{\partial L}{\partial x(k)_n}.\label{eq:LB}
\end{align}

\subsection{Truncated Backpropagation}\label{subsec:bp_simplified}

For $(\mathrm{bpv})$ in Eq.~(\ref{eq:bpval}), the value of $\frac{\partial L}{\partial \boldsymbol{r}}$ is determined at time $T$,
so the reservoir state must be stored for the number of times used for backpropagation plus one time.
Therefore, in order to use the entire time series, $(T+1)$ reservoir states must be stored, which consumes quadratically larger memory space as $T$ becomes larger.
To alleviate the memory usage while maintaining optimization accuracy, we propose a {\it truncated backpropagation} method
where only a limited number of reservoir states, e.g., the last time only, are used.
This approximation is based on the observation that the last reservoir state cumulatively reflects past reservoir states,
and the impact of past states gradually attenuates.
With this approximation, Eqs.~(\ref{eq:bpval}),(\ref{eq:Lx}),(\ref{eq:LA}), and (\ref{eq:LB}) are respectively simplified as follows:
\begin{align}
(\mathrm{bp\ value})&=\sum_{j=1}^{N_x}x(T-1)_j\frac{\partial L}{\partial r_{(n-1)N_x+j}}+\frac{\partial L}{\partial r_{N_x^2+n}},\\
\frac{\partial L}{\partial x(T)_n}&=(\mathrm{bp\ value})+B\frac{\partial L}{\partial x(T)_{n+1}},\\
\frac{\partial L}{\partial A}&=\frac{\partial f(j(T)_n+x(T-1)_n)}{\partial A}\frac{\partial L}{\partial x(T)_n},\label{eq:LA2}\\
\frac{\partial L}{\partial B}&=x(T)_{n-1}\frac{\partial L}{\partial x(T)_n}.\label{eq:LB2}
\end{align}
By comparing Eqs.~(\ref{eq:LA}),(\ref{eq:LB}) and Eqs.~(\ref{eq:LA2}),(\ref{eq:LB2}),
it becomes evident that the computational effort is significantly reduced by about $1/T$.
This simplification allows us to store only two reservoir states, i.e., $\boldsymbol{x}(T-1)$ and $\boldsymbol{x}(T)$.
Specifically, for many datasets with a time series length $T$ greater than 100,
the memory requirement for the reservoir state can be decreased to less than 2\%.
Considering the overall memory usage, which includes output weights and other factors,
let us consider a scenario that a DFR solves a three-class classification task with a time series length of $T=500$ and $N_x=30$,
the reduction in memory usage would be approximately 80\%.

\section{Evaluation}\label{sec:evaluation}

In this section, we evaluate the proposed parameter optimization method for DFR using the same datasets (npz files) as in~\cite{bianchi2020reservoir}.
Throughout the evaluation, 
the optimization of the reservoir function $f$ in modular DFR was not performed;
$f(x)=Ax$ was used consistently for all datasets, as suggested in~\cite{ikeda2023modular}.
Also, the reservoir size, $N_x$, is set to 30.
The evaluation was performed in the following environment:
AMD Ryzen 7 5825U CPU with 16.0~GB of memory.
All programs were written using Python 3.9.6 and the numpy library version 1.23.0.

In the proposed method,
The reservoir parameters, $A$ and $B$, and the output parameters, $W_{\mathrm{out}}$ and $b$, are optimized by backpropagation.
The initial value of $[A, B]$ is $[0.01,0.01]$, and the output parameters are initialized to zeros.
The stochastic gradient descent method was run over 25 epochs.
The learning rate starts at 1 and is multiplied by 0.1
for the parameters of the reservoir layer at epochs 5, 10, 15, and 20.
For the parameters of the output layer, the learning rate is multiplied by 0.1 at epochs 10, 15, and 20.
Once the optimization with backpropagation is completed, the output layer undergoes training using ridge regression with the regularization parameter $\beta$.
For $\beta$, we try $[10^{-6}, 10^{-4}, 10^{-2}, 10^{0}]$ and choose the one with the smallest loss $L$.

\subsection{Comparison with grid search}

First, we compared the proposed method with grid search,
executed in three dimensions, $A$, $B$, and $\beta$.
The ranges for $A$ and $B$ were set to $[10^{-3.75},10^{-0.25}]$, $[10^{-2.75},10^{-0.25}]$, respectively.
These ranges were determined to be able to find the optimal parameters for all the datasets used in this evaluation.
In these ranges, the number of grid divisions is increased from 1 until the accuracy reaches that of the proposed method.
Note that the grid divisions are performed equally,
e.g., if the number is 2, the range is divided into two equal sections.
$\beta$ is searched in the same way as the proposed method.

\begin{figure}[tb]
 \centering
 \includegraphics[width=0.8\linewidth]{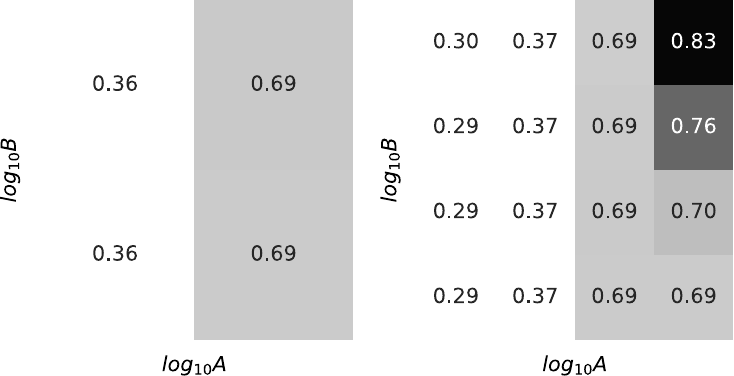}
 \caption{Grid search example where the recursively narrowing the finer grids is difficult. The dataset used is CHAR. Left: grid level 1, right: grid level 2.}\label{fig:grid}
 \vspace{-5mm}
\end{figure}

The comparison results for accuracy and computation time are summarized in Table~\ref{tab:vsgrid}.
Remarkably, the accuracy achieved by the parameters optimized with backpropagation is equal to or even superior to the state-of-the-art study~\cite{ikeda2022hardware} for most datasets.
\begin{table}
 \centering
 \caption{Runtime comparison between the proposed backpropagation (bp) and grid search (gs).
 Column ``gs divs'' represents the number of grid divisions required to achive the equal accuracy as the proposed method (bp acc).
}
 \label{tab:vsgrid}
 \begin{tabular}{c|ccccc} \hline
  data & bp & bp & gs & gs & (gs time)/ \\
  set & acc & time (s) & divs & time (s) & (bp time) \\ \hline \hline
  ARAB & 0.981 & 245 & 8 & 25,040 & 102.2 \\ \hline
  AUS & 0.954 & 54 & 8 & 5,535 & 102.5 \\ \hline
  CHAR & 0.918 & 44 & 10 & 4,820 & 109.5 \\ \hline
  CMU & 0.931 & 4 & 1 & 3 & 0.8 \\ \hline
  ECG & 0.850 & 11 & 16 & 4,977 & 452.5 \\ \hline
  JPVOW & 0.978 & 4 & 4 & 106 & 26.5 \\ \hline
  KICK & 0.800 & 7 & 1 & 2 & 0.3 \\ \hline
  LIB & 0.806 & 12 & 18 & 8,423 & 701.9 \\ \hline
  NET & 0.783 & 45 & 1 & 49 & 1.1 \\ \hline
  UWAV & 0.850 & 65 & 10 & 6,322 & 97.3 \\ \hline
  WAF & 0.983 & 14 & 3 & 188 & 13.4 \\ \hline
  WALK & 1.000 & 4 & 1 & 3 & 0.8 \\ \hline
 \end{tabular}
\end{table}
The proposed method significantly reduces the computation time, up to approximately 1/700th of the grid search.
This reduction is particularly significant for datasets where the grid search requires long computation times.
For datasets requiring fewer grid divisions, the parameter space yielding good results is larger, or the initial coarse grid produces satisfactory outcomes coincidentally.
However, early stopping of grid search is practically challenging, as it is difficult to know whether optimal accuracy has been attained or not.
On the other hand, datasets with a small area of optimal parameters require a larger number of grid divisions for accurate optimization.
%
Grid search, in the worst case, demands exponential computation time.
An alternative method to explore a finer parameter space is to recursively dig the best region in the coarser search,
which results in linear time complexity.
However, this approach may fail for certain datasets if the coarse grid search does not yield globally optimal parameters,
as illustrated in the example shown in Fig.~\ref{fig:grid}.
In contrast, the proposed method successfully found optimal values with a fixed number of epochs for all datasets used in this study.
Although further improvements can be explored by attempting different initial values or increasing the number of epochs, these did not improve the accuracy of the presented datasets.

\subsection{Memory usage reduction by {\sato{truncated}} backpropagation}

Table~\ref{tab:comp3} presents the memory savings resulting from the truncation of the backpropagation calculations shown in Section~\ref{sec:propose}.\ref{subsec:bp_simplified}.
\begin{table}
 \centering
 \caption{Storage reduction by truncated backpropagation.
 ``Naive'' and ``simplified'' represent the total number of stored values of the reservoir state, reservoir representation, and reservoir weight,
 before and after truncating the computation of backpropagation.
 }\label{tab:comp3}
 \begin{tabular}{c|ccc} \hline
  dataset & naive (a) & simplified (b) & (a-b)/a \\ \hline\hline
  ARAB & 13030 & 10300 & 21~\% \\ \hline
  AUS & 93455 & 89435 & 4~\% \\ \hline
  CHAR & 25700 & 19610 & 24~\% \\ \hline
  CMU & 20192 & 2852 & 86~\% \\ \hline
  ECG & 7352 & 2852 & 61~\% \\ \hline
  JPVOW & 10179 & 9369 & 8~\% \\ \hline
  KICK & 28022 & 2852 & 90~\% \\ \hline
  LIB & 16245 & 14955 & 8~\% \\ \hline
  NET & 42853 & 13093 & 69~\% \\ \hline
  UWAV & 17828 & 8438 & 53~\% \\ \hline
  WAF & 8732 & 2852 & 67~\% \\ \hline
  WALK & 60332 & 2852 & 95~\% \\ \hline
 \end{tabular}
\end{table}
While the reduction in memory is relatively smaller for datasets with a large number of classes and short series lengths,
more than half of the data sets still achieve a reduction larger than 50\%.
Additionally, for all datasets, including those with small memory reductions,
the computational effort of backpropagation has been reduced to about $1/T$,
resulting in a considerable reduction in computation time.

\section{Conclusion}
\label{sec:conclusion}

We proposed a fast and accurate parameter optimization method for DFR utilizing backpropagation and gradient descent.
Back propagation in the reservoir layer, which was particularly challenging and computationally intensive, was addressed by introducing the modular DFR model.
Using the model, we formulated the backpropagation of DPRR, which is effective in improving accuracy but poses difficulties due to long dependence on past reservoir states.
Additionally, we presented a truncated backpropagation method that takes advantage of reservoir state's characteristics.
As a result, the proposed method significantly reduced the computation time by up to about 1/700.
Moreover, the truncation technique successfully reduces overall memory usage by half for most datasets.


\bibliographystyle{ACM-Reference-Format}
\bibliography{ref}

\end{document}